\documentclass[12pt,a4paper]{article}
\usepackage{amssymb,amsfonts,amsmath}
\usepackage{color}
\usepackage{epsfig, euscript, graphics}

\begin{document}

\title{Getting rid of nonlocality from quantum physics}

\author{Andrei Khrennikov\\ 
Linnaeus University, International Center for Mathematical Modeling\\  in Physics and Cognitive Sciences
 V\"axj\"o, SE-351 95, Sweden}

\maketitle

\abstract{This paper is aimed to dissociate nonlocality from quantum theory. We demonstrate that the tests on violation of the Bell type inequalities are simply statistical tests of local incompatibility of observables. In fact, these are tests on violation of {\it the Bohr complementarity principle.} Thus,   the attempts to couple  experimental violations of the  Bell type inequalities with ``quantum nonlocality'' is 
really misleading. These violations are explained in the quantum theory as exhibitions of incompatibility of 
observables for a single quantum system, e.g., the spin projections for a single electron or the polarization projections for a single photon. Of course, one can  go beyond quantum theory with the hidden variables models (as was suggested by Bell) and then discuss their  possible nonlocal features.  However, conventional quantum theory is local.}

\section{Introduction}

As is well known, the~original EPR-argument~\cite{EPR} was fundamentally coupled with 
the Bohr complementarity principle~\cite{BR0,PL1,PL2} (see Section~\ref{COMPP}). Einstein, Podolsky, and~Rosen  reasoned against the  completeness of quantum mechanics (QM) by showing that the ``elements of reality'' corresponding to two incompatible observables (e.g., position and momentum) can be assigned  to the same physical system. However, their argument was purely theoretical  (even merely philosophical) and it was impossible to check the EPR-statement experimentally. Bohr pointed to  the latter in his reply to Einstein~\cite{BR}; he considered the EPR-argument as metaphysical. Although~nonlocality  was mentioned in the EPR-paper, 
it was considered as just a possible alternative to incompleteness of~QM.

Nonlocality was emphasized for the first time by by Bohm.  It was elevated by Bell (who admired Bohmian mechanics) through the argument based on violation of Bell's inequality~\cite{Bell0,Bell1,Bell2}. 

Our aim is to perform the {\it genuine quantum mechanical analysis of the derivation of the CHSH-inequality} considered as an inequality 
for correlations of quantum observables---{\it the quantum CHSH-inequality}. Thus,    we do not try to go beyond QM. We are interested in the quantum mechanical interpretation of experimental  violation of  the CHSH-inequality.   We show that, in~fact, the~degree of violation is straightforwardly coupled to the degree of incompatibility of observables,  the~norms of commutators.  In~particular, in~the scenario  with two spatially separated systems, these are tests of local incompatibility, i.e.,~{\it incompatibility of observables on a single subsystem of a compound system} (Theorem 1, Section~\ref{LR}). 

We remark that the CHSH-inequality~\cite{CHSH} is derived for classical correlations expressed in the framework of hidden variables
(established by Bell \cite{Bell0,Bell1,Bell2}) and by using the calculus of classical probabilities (the Kolmogorov probability theory). The~quantum CHSH-inequality is derived for quantum correlations  by using the operator~formalism.

We stress that the CHSH combination of correlations,
\begin{equation}
\label{LC}
\langle  {\cal B} \rangle  =\frac{1}{2} [\langle A_1 B_1  \rangle + \langle A_1 B_2 \rangle + \langle A_2  B_1  \rangle- \langle A_2 B_2 \rangle],
\end{equation}
has three different~interpretations:
\begin{itemize}
\item Classical (hidden variables) correlations, $\langle  {\cal B} \rangle_{\rm{CL}}.$
\item Experimental correlations, $\langle  {\cal B} \rangle_{\rm{EXP}}.$
\item Quantum mechanical correlations, $\langle  {\cal B} \rangle_{\rm{QM}}.$
\end{itemize}

Following Bell,  one can compare the classical theoretical quantity 
$\langle  {\cal B} \rangle_{\rm{CL}}$ with its experimental counterpart $\langle  {\cal B} \rangle_{\rm{EXP}}.$
 The majority of the quantum foundation and information community proceeds in this way. This way leads to operating with the notion of nonlocality and action at a~distance. 

{\it Is there any reason to couple this mysticism  with quantum theory?}

It is more natural to start with the quantum theoretical analysis of quantity $\langle  {\cal B} \rangle_{\rm{QM}}.$ It is easy to explain 
under what circumstances it can be bounded  by 1 (or exceed 1).\footnote{Thus so-called classical bound has the purely quantum origin.}  This quantum mechanical explanation is purely local. Thus, it is really incorrect in quantum theory to speak about its nonlocality or associate with it any kind of action at a~distance. 

It is well known that (by the complementarity principle) it is impossible to measure 
jointly two spin coordinates for electron. Therefore, 
$\langle  {\cal B} \rangle_{\rm{QM}}$ can exceed 1. If~somebody does not believe  in this prediction of QM,
it would be natural to check violation of the principle of complementarity for a single electron (or photon),
e.g., to~check violation of the Heisenberg uncertainty relation (in the form of the Robertson inequality).

As  emphasized above, here I  proceed by using solely the formalism of QM, cf. with probabilistic analysis of the incompatibility interpretation of the Bell type {inequalities in}  \cite{Accardi}-\cite{BC2}
and especially~\cite{KHB2} (the  probabilistic version of the present paper). (See also 
the recent preprint of Griffiths~\cite{Griffiths}, where incompatibility  of quantum observables is emphasized; see the recent works of Boughn~\cite{Boughn1,Boughn2}, where the nonlocality viewpoint on quantum theory is critically analyzed and 
the role of the ontological vs. information interpretations of the wave function in discussions on ``quantum nonlocality'' is~emphasized.) 

Foundational issues such as the complementarity principle, incompatibility, nonlocality, realism, and~hidden variables,    are  discussed in more detail in Section~\ref{FOOP}. 

\section{Measuring the Degree of Incompatibility via the~CHSH-Test}

We   show that the degree of violation the quantum CHSH-inequality can be considered as a measure of incompatibility in two pairs of quantum observables, $A_1,A_2$ and $B_1, B_2.$ This is the simple consequence of the   {\it the Landau identity}  \cite{Landau,Landau1}  (see Equation  \eqref{L2}). In~quantum theory incompatibility is mathematically expressed as noncommutativity. Thus,    by~testing incompatibility, we  test 
the degree of noncommutativity, or~in other words, the~``magnitudes" of observables corresponding to commutators,
\begin{equation}
\label{INC}
\hat M_A=i [\hat A_1, \hat A_2], \;  \hat M_B=i [\hat B_1, \hat B_2].
\end{equation}

We  use the hat-symbol  to denote~operators.

The incompatibility-magnitude can be  expressed via the maximal value of averages of commutator-operators, i.e.,~by their norms, for~example,
\begin{equation}
\label{MG}
\sup_{\Vert \psi\Vert=1} \vert\langle \psi\vert \hat M_A\vert \psi\rangle\vert = \Vert \hat M_A\Vert.
\end{equation}

By interpreting quantity $\langle \psi\vert \hat M_A\vert \psi\rangle$ as the 
theoretical counterpart 
of experimental average $\langle M_A\rangle_\psi$ of observable $M_A,$ we can measure  experimentally  the incompatibility-magnitude, i.e.,~norm  $\Vert \hat M_A\Vert$ 
from measurements of commutator-observable $M_A.$  (The main 
foundational problem is that measurement of such commutator-observables is challenging. Recently some progress was demonstrated on the basis of weak measurements, but~generally we are not able to measure 
commutator-quantities.)  

We remark that (from the quantum mechanical viewpoint) the CHSH-test estimates the product of incompatibility-magnitudes for  the $A$-observables and $B$-observables, i.e.,~the quantity $\Vert \hat M_A\Vert\Vert \hat M_B\Vert.$ However, by~considering 
the $B$-observables as axillary and selecting them in a proper way (for example, such that the $B$-commutator is a simple operator), we can use the CHSH-test to 
get the experimental value for  the incompatibility-magnitude $\Vert \hat M_A\Vert.$

\section{Incompatibility as  Necessary Condition of Violation of Quantum~CHSH-Inequality} 
\unskip
\subsection{General Case: Without Referring to the Tensor Product~Structure}
\label{GC}

{Consider} the Bohm--Bell  type experiments.  Four observables $A_1, A_2, B_1, B_2$ taking values $\pm 1$    are considered.
It is assumed that  observables in each pair $A_i, B_j, i, j=1,2,$  can be measured jointly, i.e.,~$A$-observables are compatible with $B$-observables. 
However, the~observables in pairs $A_1, A_2$ and $B_1,B_2$ are incompatible, i.e.,~they cannot be jointly measured. Thus,    probability 
distributions $p_{A_i B_j}$ are well defined theoretically in QM and they can be verified experimentally; 
probability distributions $p_{A_1 A_2}$ and $p_{B_1 B_2}$ are not defined in QM and, 
hence, the~question of their experimental verification does not~arise.  

We consider quantum observables represented by Hermitian operators. In~QM, mathematical compatibility is represented by commutativity of operators, i.e.,~in the Bohm--Bell  type experiments
\begin{equation}
\label{KO1}
[\hat A_i, \hat B_j]=0, i,j=1,2,
\end{equation}
and generally
$
[\hat A_1, \hat A_2]\not=0, \; [\hat B_1, \hat B_2]\not=0.
$
 The  quantum theoretical CHSH-correlation function has \mbox{the form:}
\begin{equation}
\label{LC}
\langle  {\cal B} \rangle   =\frac{1}{2} [\langle \hat A_1 \hat B_1  \rangle + \langle \hat A_1 \hat B_2 \rangle + \langle \hat A_2  \hat B_1  \rangle- \langle \hat A_2 \hat B_2 \rangle].
\end{equation}
(Here and everywhere below the index $\rm{QM}$ pointing to the quantum formalism is omitted.)

It is compared with the experimental CHSH-correlation~function.  

In the quantum framework,  the~CHSH-correlation function can be expressed with the aid of \mbox{the Bell-operator:}
\begin{equation}
\label{L1}
\hat {\cal B} = \frac{1}{2}[\hat A_1(\hat B_1+ \hat B_2) +\hat A_2(\hat B_1-\hat B_2)]
\end{equation}
as
\begin{equation}
\label{L1T}
\langle  {\cal B} \rangle = \langle \psi\vert \hat {\cal B} \vert \psi\rangle. 
\end{equation}

By straightforward calculation, one can derive at the Landau identity:
\begin{equation}
\label{L2}
\hat{{\cal B}}^2=I - (1/4) [\hat A_1, \hat A_2][\hat B_1,\hat B_2].
\end{equation}

Thus,    if~{\it at least one of the commutators} equals to zero, i.e.,
\begin{equation}
\label{L3}
[\hat A_1,\hat A_2]=0, 
\end{equation}
or
\begin{equation}
\label{L4}
[\hat B_1,\hat B_2]=0,
\end{equation}
then the  following inequality holds:
\begin{equation}
\label{L1n}
\vert \langle  {\cal B} \rangle  \vert   \leq 1.
\end{equation}

To derive this inequality, we used solely the quantum formalism. The~inequality is the consequence 
of compatibility for at least one pair of observables, $A_1, A_2$ or $B_1, B_2.$ Thus, although~formally 
Equation~(\ref{L1n}) coincides with the standard CHSH-inequality, it has totally different meaning. 
\mbox{It is}  better to call  Equation~(\ref{L1n}) {\it the quantum CHSH inequality.}

Thus,    {\it compatibility of the $A$-observables or the $B$-observables is sufficient  for validity of the quantum CHSH-inequality} (for all quantum states) or in other words {\it conjunction of incompatibilities of the $A$-observables and the $B$-observables  is the  necessary condition for its violation} (for some quantum state).

\subsection{Compound~Systems} 
\label{COMS}

States of a compound quantum system $S=(S_A,S_B)$ are represented  in  tensor product $H_{AB} =H_A \otimes H_B$ of the state spaces
$H_A$ and $H_B$ of subsystems $S_A$ and $S_B.$ Observables  are given by operators
\begin{equation}
\label{KO2}
\hat A_i=\hat {\bf A}_i  \otimes I, \; \hat B_i = I \otimes \hat {\bf B}_i,
\end{equation}
where Hermitian operators $\hat {\bf A}_i$ and $\hat {\bf B}_i$ act in $H_A$ and $H_B,$ respectively.
They represent observables ${\bf A}_i, {\bf B}_i$ on subsystems $S_A, S_B$ of $S.$ For spatially separated systems,
we call them {\it  local observables}.
 
This tensor representation automatically implies commutativity of operators $\hat A_i$ with operators 
$\hat B_j,$ i.e.,~ Equation~(\ref{KO1}) holds. We remark that the mathematical condition of incompatibility 
is reduced to condition 
$
[\hat {\bf A}_1, \hat {\bf A}_2]\not=0 \; \mbox{and}  \; [\hat {\bf B}_1, \hat {\bf B}_2] \not=0.
$
For spatially separated systems, it is natural to call  incompatibility of the observables on $S_A$ (on $S_B$)
  {\it local incompatibility}. 
Section~\ref{GC} implies that  {\it conjunction of local incompatibilities}
is the necessary condition for violation of the quantum~CHSH-inequality.

We remark that the mathematical formalism of this section is applicable to description of any kind 
of observables ``respecting'' the tensor product structure of observables. A~physical system $S$ need not be composed of two physical~subsystems.     

\section{Incompatibility as  Sufficient Condition of Violation of Quantum~CHSH-Inequality}
\unskip

\subsection{General Case: Without Referring to the Tensor Product~Structure}
\label{GC1}

Assume that $A$-observables as well as $B$-observables are incompatible, i.e.,~corresponding operators do not commute:
\begin{equation}
\label{L3z}
[\hat A_1,\hat A_2]\not=0 \; \mbox{and}  \; [\hat B_1,\hat B_2] \not=0,
\end{equation}
i.e.,
\begin{equation}
\label{L2zT}
\hat M_A\not=0 \;\mbox{and} \; \hat M_B\not=0,
\end{equation}
where $\hat M_A= i[\hat A_1, \hat A_2], \;  \hat M_B = i [\hat B_1,\hat B_2].$ It is important to note that 
 $
[\hat M_A, \hat M_B]=0.
$
We can write the Landau identity in the form
\begin{equation}
\label{L2z}
\hat{{\cal B}}^2=I + (1/4)\hat M_{AB},
\end{equation}
where $\hat M_{AB}= \hat M_A \hat M_B.$
If $M_{AB} =0,$ then,   despite  the incompatibility condition in Equation~(\ref{L3z}), the~QCHSH-inequality cannot be violated. We proceed under condition
\begin{equation}
\label{L2zz}
\hat M_{AB}\not=0.
\end{equation}

In our framework, this condition is not so restrictive. We consider the quantum CHSH-inequality as a statistical test 
of incompatibility. It is natural to estimate the degree of incompatibility in one pair of observables, e.g.,~
in the $A$-pair. In~this approach, the~$B$-pair plays the axillary role and we can freely play with its selection. To~obtain the condition in Equation~(\ref{L2zz}), it is sufficient to select $B$-operators in such a way 
that the operator    $\hat M_{B}$  is invertable. We especially highlight the case of compound systems (see Section~\ref{COMS}). Here  incompatibility of the $A$-observables and the $B$-observables, see Equation~(\ref{L2zT}), automatically implies the condition in Equation~(\ref{L2zz})

Under the condition in Equation~(\ref{L2zz}), there exists some common eigenvector $\psi_{AB}$ such that $M_A \psi_{AB}= \mu_A \psi_{AB},
M_B \psi_{AB}= \mu_B \psi_{AB}$ and both eigenvalues are~nonzero. 

Suppose that $\mu_A>0$ and  $\mu_B >0.$ Then,   this $\psi_{AB}$ is an eigenvector of operator $\hat{\cal B}^2$ with 
eigenvalue $(1+\mu)>1, \mu=\mu_A \mu_B.$ Hence, $\Vert \hat {\cal B}^2 \Vert \geq (1+ \mu)>1$ and 
$$     
1< (1+ \mu) \leq \Vert \hat {\cal B}^2 \Vert = \Vert \hat {\cal B} \Vert^2.
$$

Since $\hat {\cal B}$ is Hermitian, we have
$$
\Vert \hat {\cal B}\Vert = \sup_{\Vert \psi \Vert =1} \vert \langle \psi\vert \hat {\cal B} \vert \psi\rangle \vert.
$$

Finally, we get that 
$$
\sup_{\Vert \psi \Vert =1} \vert \langle \psi\vert \hat {\cal B} \vert \psi\rangle \vert > \sqrt{1+ \mu} >1.
$$

Thus,    there exist pure quantum states such that the QCHSH-inequality is~violated.

Now, suppose that $\mu_A >0,$  but  $\mu_B < 0.$ To change the sign of   $\mu_B,$ it is sufficient 
to interchange the $B$-observables.

Thus,    {\it conjunction of  incompatibilities of  the $A$-observables and the $B$-observables  constrained by Equation~(\ref{L2zz}) is sufficient for violation of the quantum CHSH-inequality.}

\subsection{Compound~Systems}
\label{LR}

Here, $H=H_A\otimes H_B$ and $\hat A_j=  \hat {\bf A}_j \otimes I, \hat B_j=  I \otimes \hat {\bf B}_j,$
where Hermitian operators   $\hat {\bf A}_j$ and $\hat {\bf B}_j$ act in $H_A$ and $H_B,$ respectively. 

\subsubsection{Incompatibility as Necessary and Sufficient Condition of Violation of the Quantum~CHSH-Inequality} 

Here, the~joint incompatibility-condition in Equation~(\ref{L3z}) is equivalent to incompatibility of observables on subsystems:
\begin{equation}
\label{L3za}
\hat {\bf M}_A= i [\hat  {\bf  A}_1,\hat {\bf  A}_2]\not=0 \; \mbox{and}  \; \hat {\bf M}_B= i [\hat {\bf  B}_1,\hat {\bf B}_2] \not=0. 
\end{equation}

We have $\hat M_{AB}= \hat M_A \hat M_B= \hat {\bf M}_A \otimes \hat {\bf M}_B.$ 
As     mentioned above, constraint $\hat M_{AB}\not=0$ is equivalent to Equation~(\ref{L3za}). 
Section~\ref{GC} implies that  {\it conjunction of local incompatibilities}
is the sufficient condition for violation of the quantum CHSH-inequality. Thus, we obtain: 

\medskip 

{\bf Theorem 1} [Local incompatibility criteria of QCHSH-violation]
 {\it Conjunction of local incompatibilities 
is the necessary and sufficient condition for violation of the quantum CHSH-inequality.}

\subsubsection{Eigenvectors of the Bell Operator and Its~Square} 
\label{EV}
  
Consider the eigenvector consideration of Section~\ref{GC1}. The~vector $\psi_{AB}= \psi_{A} \otimes \psi_{B},$
where $\psi_{A} \in H_A, \psi_{B} \in H_B,$ and $\hat {\bf M}_A\psi_{A} = \mu_A \psi_{A},
 \hat {\bf M}_B \psi_{B} =\mu_B \psi_{B}.$ We assume that $\mu=\mu_A\mu_B >0.$ Then,
$$
\langle \psi_{A} \otimes \psi_{B} \vert \hat {\cal B}^2 \vert  \psi_{A} \otimes \psi_{B} \rangle  > 1.
$$ 

Thus, for~the squared CHSH-observable $\hat {\cal B}^2,$ the one-boundary can be violated for separable states.
Here, entanglement of $A$ and $B$ observables plays no~role.  

To be more illustrative, let us restrict consideration to finite dimensional Hilbert spaces.
There can be found states $\Psi$ and $\Phi$ such that
$$
\max_{\Vert \psi \Vert=1} \vert \langle \psi \vert \hat {\cal B}^2 \vert  \psi  \rangle\vert = 
 \langle \Psi \vert \hat {\cal B}^2 \vert  \Psi \rangle,
$$ 
$$
\max_{\Vert \psi \Vert=1} \vert \langle \psi \vert \hat {\cal B} \vert  \psi \rangle=
\vert \langle \Phi \vert \hat {\cal B} \vert  \Phi\rangle.
$$ 

The tricky thing is that generally $\Psi \not=\Phi.$ The equality for norms, $\Vert \hat {\cal B} \Vert= 
\sqrt{\Vert \hat {\cal B}^2 \Vert},$ does not imply equality of the  $\max$-optimization~states.

Of course, $\max$-states for ${\cal B}$ and ${\cal B}^2$ are connected: the former can be represented as linear combinations of the latter 
(the feature of all operators with degenerate spectrum). (As   shown 
in \cite{Braunstein},  $\max$-states for ${\cal B}$ can be represented even as mixtures of $\max$-states for ${\cal B}^2$.)

In the experiments to violate the quantum CHSH-inequality, tremendous efforts were put to prepare ensembles of 
entangled states. The~main reason for this is that the direct measurement of the observable represented by operator $\hat {\cal B}^2$ is challenging. In~{Appendix \ref{appb}}, we present the abstract analog of the Bell experiments
treated as experiments to measure the degree of incompatibility. The~tensor product structure is excluded and, in~particular, an~analog of entangled states related to measurement of an observable and its square is~considered. 

\section{CHSH-Correlation Function as Measure of~Incompatibility}

We start with consideration of  observables respecting  the tensor product structure on the state space 
$H=H_A\otimes H_B.$
Consider the eigenbases $(e_{Ak})$ and  $(e_{Bk})$ of operators $\hat {\bf M}_A$ and $\hat {\bf M}_B$ 
(acting in $H_A$ and  $H_B,$ respectively) and the corresponding 
eigenvalues $\mu_{Aj}, \mu_{Bj}.$ 

Let $\Vert \hat {\bf M}_A\Vert = \max_j \vert \mu_{Aj}\vert= \vert \mu_{Ai_a} \vert,
\Vert \hat {\bf M}_B\Vert =\max_j \vert \mu_{Bj}\vert= \vert \mu_{Bi_b} \vert$ and let 
$\mu_{Ai_a} \mu_{Bi_b} >0.$
Then,   $\Vert \hat {\cal B}^2 \Vert= 
(1+ \mu_{Ai_a}\mu_{Bi_b}).$ Thus,
\begin{equation}
\label{IV}
b= \Vert \hat {\cal B} \Vert = \sqrt{1 + \frac{1}{4} 
\Vert [\hat {\bf A}_1, \hat {\bf A}_2]\Vert \; \Vert [\hat {\bf B}_1, \hat {\bf B}_2]\Vert},
\end{equation}  
where $\langle {\cal B}\rangle_\psi$ is given by Equation~(\ref{LC}); $b$  is the maximal 
possible value of CHSH-correlations. If~eigenvalues $\mu_{Ai_a}$ and $\mu_{Bi_b}$ have different  signs, then
we interchange the $B$-observables. 

From Equation~(\ref{IV}), we get that 
$$
\Vert [\hat {\bf A}_1, \hat {\bf A}_2]\Vert \Vert [\hat {\bf B}_1, \hat {\bf B}_2]\Vert= 4 (b^2-1).
$$

The norm of commutator can be considered as a measure of incompatibility.
Thus,    \mbox{the CHSH-} correlation function gives the possibility to check experimentally 
the product of degrees of incompatibility for the $A$ and $B$ observables.

One may consider this way of measuring of incompatibility as too tricky. \mbox{However, typically,} to~measure 
commutator-observable and then its average
is challenging . (By ``measuring 
commutator- observable'', we mean measuring observable represented mathematically by commutator operator scaled by $i.$)  Therefore, even such
a tricky approach to this problem as measurement of the CHSH-correlation function deserves 
attention.

Now, consider $B$-observables as axillary. In~this way, we are able to determine the degree of incompatibility for 
the $A$-observables by using some pair of axillary observables $B_1, B_2.$ We can select the latter in such a a way that their commutator is a ``good observable'', so that it can be easily measured for any state, 
thus its average and hence the norm can be determined. Then,   we can measure incompatibility  of 
observables $A_1$ and $A_2$ by using the formula:
\begin{equation}
\label{qqq}
\Vert [\hat {\bf A}_1, \hat {\bf A}_2]\Vert = 4 (b^2-1)/ \Vert [\hat {\bf B}_1, \hat {\bf B}_2]\Vert .
\end{equation} 

Why is the use of tensor product states  so useful for measuring the degree of incompatibility?  
\mbox{By spitting} a system into two subsystems it is easy to check compatibility of $A$ and $B$ observables, 
\mbox{thus the} possibility to define the CHSH-correlation function which can be measured in~experiment. 

\section{Foundational~Questions}\label{FOOP}
\vspace{-6pt}

\subsection{Bohr's Complementarity~Principle}
\label{COMPP}

Often, it is claimed that  Bohr's writings and, in~particular, about the complementarity principle
are very difficult for understanding. (For example, 
in Schilpp's volume~\cite{Schilpp}, p.~674 (see~\mbox{also Plotnitsky~\cite{PL3}}, p.~108),  one can find the following statement: ``{\it Thus, Einstein was confessed, after~decades of his exchanges with Bohr, that he was `unable to attain ... the sharp formulation ... [of] Bohr's principle of complementarity'''.}) 
This principle has the complex structure and composed of a few components. 
One of the problems is that typically this principle is reduced to just one of its components, namely,
the incompatibility-component. Incompatibility has the most striking consequences for quantum 
theory and experiment.  However, as separated from the body of the complementarity principle, 
incompatibility is difficult for~understanding. 

As   emphasized in~\cite{KHB2}, the~complementarity principle is in fact the principle of contextuality 
of quantities used in the quantum formalism, in~the sense of coupling them to 
corresponding experimental contexts.  Bohr did not use 
the notion  ``experimental context". He considered experimental
conditions~\cite{BR0}:

{\it ``Strictly speaking, the~mathematical formalism of quantum mechanics
and electrodynamics merely offers rules of calculation for the deduction of
expectations pertaining to observations obtained under well-defined experimental 
conditions specified by classical physical concepts.''} 

By using the notion of experimental context as the synonymous  of Bohr's  experimental conditions we can present the complementarity principle as composed 
of the following components~\cite{KHB2}. (We remark that one has to be very careful by operating with the notion of contextuality.
Nowadays, this notion is widely used in foundational 
discussion on the Bell type inequalities. In~such discussions, the~meaning 
of the notion contextuality does not coincide with 
Bohr's contextuality, as~taking into account the experimental context to explain the mechanism of generating the values of a quantum
observable. From~Bohr's viewpoint, any single quantum observable is
contextual. One may say that consideration of Bohr's contextuality in parallel 
with Bell's contextuality can be misleading. However, we can consider Bell's contextuality 
simply as a very special case of Bohr's contextuality.)
  
\begin{itemize}
\item (B1): There exists the fundamental quantum of action given by
the Planck constant $h$:
\item (B2): The presence of $h$ prevents approaching internal features
of quantum systems.
\item (B3): Therefore, it is meaningless (from the viewpoint of physics)
to build scientific theories about such features.
\item(B4): An output of any observable is composed of contributions
from a system under measurement and the measurement device.
\item (B5): Therefore, the~complete experimental arrangement
 (context) has to be taken into account.
\item (B6): There is no reason to expect that all experimental contexts
can be combined. Therefore, \mbox{there is} no reason to expect that all
observables can be measured jointly.  Hence, there exist incompatible
observables.
\end{itemize}

(B6) can be called the incompatibility principle; this is a consequence of (B4)
and (B5). Typically, the~complementarity principle is identified with (B6). However,
such a viewpoint does not match Bohr's understanding of the complementarity principle,
as the combination (B1)--(B6).    

This  is the good place to remark that  (B6) is very natural. The~existence of incompatible experimental contexts is not surprising. Compatibility of all experimental contexts
would be really~surprising.

\subsection{``Quantum~Nonlocality''}

We briefly discuss the notion of (non)locality. 

\subsubsection{Relativistic~Invariance}

Everywhere in physics, besides~the Bell inequality debates~\cite{Bell0,Bell1,Bell2,CHSH,Braunstein,Hardy,Cereceda,Wolf,Mermin}, locality is identified with the relativistic invariance of theory. Therefore, the~statements on nonlocality of quantum theory can make the impression 
(and they do!) that there is something wrong with relativistic invariance. However, there is nothing wrong 
with relativistic invariance. Of~course, QM (in particular, the~Schr\"odinger equation)  is not relativistically invariant and attempts to construct relativistically invariant QM (based on the Dirac equation)  were not successful. However, {\it QM is an approximation of quantum field theory which is relativistically invariant}  
(see Bogolubov and Shirkov~\cite{BS} and Haag~\cite{Haag} (especially  Chapter 3, ``Algebras of Local 
Observables and Fields'')).
(To complete the picture, we~remark that there is a non-relativistic quantum field theory (see for example, book~\cite{NQFT}).)

\subsubsection{Hidden Variables and Action at a~Distance}

One can say that nonlocality is a consequence of ``action at the distance'' \cite{Bell0,Bell1,Bell2} (see, e.g., 
Shimony~\cite{Shimony,Shimony1} and Jaeger~\cite{Jaeger, Jaeger1} for the detailed presentation). This interpretation is based on the invention of hidden variables. However, the~analysis presented in this paper shows clearly  that, to~proceed in this framework, {\it one has to start with
rejection of  the basic principle of QM, the~complementarity principle.} It is not clear 
why violations of this principle should  be sought for compound systems. Thus, before~inventing hidden variables, it would be natural to find violations of say the Heisenberg uncertainty principle (in the form
of Robertson inequality).   

Moreover, the~modern attempt to go beyond QM  with hidden variables is too straightforward. Already in the 19th century, in~the process of transition from Newtonian mechanics to classical field theory, physicists
were confronted with the same problem as in the process of transition from classical physics to QM. It was resolved in the framework
of Bild (image) methodology developed by  Hertz and Boltzmann~\cite{Hertz1,Boltzmann,Boltzmann1} (see Section~\ref{HertzS} and papers~\cite{DA,Hertz,KHB2}).

\subsubsection{Nonlocality = Violation of the Bell Type~Inequalities.}

The common comment to my talks  is that per definition ``quantum nonlocality''  is a violation of the Bell type inequalities. However, this viewpoint is  really misleading. If~one recognizes that such violation is just a signature of incompatibility, then it is strange to speak about nonlocality, instead~of~complementarity.

\subsection{Obscuring Incompatibility by Tensor Product Structure of~Observables}

As   pointed out, we concentrate our analysis on the CHSH-inequality~\cite{CHSH}.
In contrast to the previous studies (see, e.g.,~\cite{Braunstein,Hardy,Cereceda,Wolf,Mermin}), we do not emphasize  the role of the tensor product structure  for the state space and observables. We proceed in the general framework and the tensor product 
model is just a special case of this framework. The~common emphasis of the tensor product 
structure obscures the crucial role played by incompatibility of observables. Mathematically 
the crucial role of incompatibility-noncommutativity  for violation of the CHSH-inequality 
was clarified  already by Landau~\cite{Landau,Landau1}  (see also~\cite{Braunstein,Hardy,Cereceda,Wolf,Mermin}). However, the~mathematical calculations presented in these works did not lead to reinterpretation of violation of the~CHSH-inequality.

I would like to emphasize the crucial role played by works of Landau~\cite{Landau,Landau1}. In~fact, Landau's 
articles carried the same message as the present paper: the CHSH inequality is an experimental test of the 
principle of complementarity. (He even used the terminology ``complementary observables'', instead of 
``incompatible observables''.) Unfortunately,  his excellent mathematical work was not completed by an extended
interpretational discussion.  Surprisingly, nowadays, his works are practically forgotten. (Of course, 
qualified people are aware about papers~\cite{Landau,Landau1}. However, generally, the~members of the quantum foundational community 
practically never refer to these papers. During~\mbox{20 years} of debates on the Bell inequality in V\"axj\"o,
 I have never heard about them. I got to know about Landau's works from E. Dzhafarov, an~expert in mathematical psychology. It happened say eight years ago and I also ignored the complementarity message of Landau.
I was content to   enjoy  his mathematics.) I see two reasons for~this:
\begin{enumerate}
\item Landau used the abstract framework of $C^\star$-algebras and, for~many ``real physicists'', this was not so attractive.
\item He emphasized the novel way to derive the Tsirelson bound and typically this paper is considered as devoted to this derivation, i.e.,~its crucial component, coupling of Bell's argument to Bohr's principle of 
complementarity was completely ignored.
\end{enumerate}     

In the present paper, I select the intermediate strategy  for representation. On~the one hand, \mbox{I do} not just follow
Landau using the language of  $C^\star$-algebras. On~the other hand, I also do not want to follow the common path 
based on the tensor product representation. I proceed in the complex Hilbert space formalism, but~generally without referring to the tensor product structure of operators. Mathematics is really simple. It is based on the interrelation of spetcral properties of the Bell operator ${\cal B}$ and its square ${\cal B}^2.$
(In fact, 
I have the impression that the essence of CHSH-test is this spectral interplay  between the spectral properties of a Hermitian operator and its square.  I try 
to present this vision in the abstract form in Appendix 2.)

\subsection{Herz--Boltzmann Bild-Methodology of~Science}
\label{HertzS}

It is surprising that  not only Bell, but even Einstein, Bohr, and Heisenberg did not know  about the works of Hertz and Boltzmann~\cite{Hertz1,Boltzmann,Boltzmann1}  on so-called ``Bild'' (image) methodology for physical theories. 
According to Hertz and Boltzmann,  when speaking about a scientific theory, one has
to specify its type: descriptive  theory or observational
theory. The~crucial point is that a descriptive theory need not be straightforwardly coupled with
theory of observations. By~extending the Hertz--Boltzmann methodology to the quantum domain, we recognize that 
QM is an observational theory. Theories with hidden variables are of the descriptive type. The~same observational 
theory can be based on a variety of descriptive theories. Bell's type descriptive theories have very rigid coupling 
to QM, the~observational theory. One can construct a variety of corresponding  descriptive theories which are not constrained
by the Bell type~inequalities. 

In this paper, we do not   to discuss the Bild-methodology of Herz and Boltzmann~\cite{Hertz1,Boltzmann,Boltzmann1} in much detail (see my recent article~\cite{Hertz}).
We only make the remark on the notion of realism. From~the Bild-viewpoint, realism in physics as well as any
other area of science is reduced solely to experimental facts. In~QM,
this is exactly Bohr's point of view. Thus,    the~only realistic component of
QM are outcomes of measurements (Bohr's phenomena).
Any physical theory (descriptive as well as observational) is only about human images of natural phenomena.
At the same time,  these images are created on the basis of human's
interaction with~nature. 

Neither Einstein nor Bohr was not aware of the works of Hertz and Bolzmann. (In any event, \mbox{they   never} 
cited these works.) Both Einstein and Bohr identified descriptive and observational theories.
In fact, the~EPR-paper~\cite{EPR} can be considered as  the message that QM is not a descriptive theory. However, at~the same time, Einstein--Podolsky--Rosen dreamed of a descriptive theory with the straightforward coupling to observations. According to Hertz and Boltzmann, the~latter 
is generally impossible. In~his reply~\cite{BR}, Bohr tried to explain that QM is an observational theory and such things as the EPR
elements of reality do not belong to its domain. However, nobody was aware about Hertz--Boltzmann distinguishing of descriptive 
and observational theories. Therefore their discussion can be compared to conversation of the blind with the~deaf.

Finally, we refer to an example of descriptive theory coupled to QM (treated as an observational theory). 
This is {\it prequantum classical statistical field theory} (PCSFT), which was developed by the author of this 
paper and coauthors~\cite{PQ} (see {Appendix 3).

\subsection{On Incompatibility Interpretation of the Bell Type~Inequalities}

	In this paper, we analyze  the CHSH-inequality and conclude  that this is a test of the complementarity 
principle. It seems that this analysis can be extended to other Bell type inequalities. The~crucial mathematical step in this analysis is derivation of the analogs of the Landau identity \mbox{(see  Hardy}~\cite{Hardy} for generalization of the
CHSH inequality to an $N$-measurement scheme and Cereceda's paper~\cite{Cereceda}, where the very general case (including Mermin's inequalities) was studied in very detail).   

In Appendix 1, we show (independently of results based on the Landau identity  that incompatibility for at least one pair of observables is the necessary condition 
for violation of any type of the Bell type inequalities.  
	
\section{Conclusions}

The point of Bell's theorem is that a local hidden variables theory cannot reproduce the results 
of quantum theory. The implication is that only a nonlocal hidden variable theory (like Bohmian mechanics) 
can reproduce the correlations found in quantum theory (and in the real world).\footnote{For the moment, we follow the conventional approach to interrelation of subquantum and quantum theories which was established by Bell
(see section \ref{HertzS} and Appendix 3 for more general picture due to Hertz and Boltzmann).}  Here, clearly, ``nonlocality'' refers to hidden variables theories, not to quantum theory.   The very common misconception is to (incorrectly) associate 
the term with quantum theory.\footnote{In particular, from~this viewpoint, the~comments of Aspect~\cite{Aspect1} and Wiseman~\cite{Wiseman} on the crucial experiments~\cite{Hensen,Giustina,Shalm} are really misleading (cf. with the comment of Kupczynski~\cite{BC2} and 
the author of this paper \cite{ABELL}).}  We hope that the argument presented in this paper has convincingly demonstrated that
this association is wrong.  The deeper message of this paper  is that the Bell 
inequality can be reinterpreted as {\it a condition on the quantum compatibility of local 
observables.}  If local commutators vanish, the correlations are bounded just as 
they are when  hidden variables  are assumed to be local.   The Bell type inequalities have one interpretation 
for hidden variables theories (the classical case), and another lesser-known and very interesting 
one for quantum theory.

Consequently the outputs of experiments testing violation of the Bell type inequalities also can be interpreted in two different ways.
The conventional interpretation is that these were classical vs. quantum physics tests. My interpretation is that such experiments were, in fact, the tests of local incompatibility of quantum observables. I claim that the latter interpretation does not diminish the foundational 
value of these breakthrough experiments \cite{Aspect,Weihs,Hensen,Giustina,Shalm}. Complementarity is the basic feature of quantum observables. Tests on this feature are of the great foundational importance. At the same time the conventional interpretation 
of these tests, local realism\footnote{This is the good place to stress that Bell has never used this notion, see the 
collection his papers in book \cite{Bell1}.} vs. quantum theory is, in fact, not so exciting. What is the meaning to test nowadays quantum against classical? The validity of quantum theory was confirmed by the huge body of experiments and technological applications. 

We also point out that the Bell type experiments  played the crucial role in development of quantum technology: creation of efficient sources of entangled systems and photo-detectors as well  as transmission of quantum systems to long distances with minimal disturbance. 

It is clear that  to get rid of nonlocality  from 
quantum theory is not a simple task. The~present note is just a step towards the common acceptance of the local interpretation of~QM.

This paper should not be considered as directed against attempts to go beyond QM, by~introducing ``hidden variables''. However, in~such 
attempts, one has to take into account the basic principles of QM an especially the complementarity principle (see the recent article 
of Khrennikov and Alodjants~\cite{KHB3}). One also has to take into account the lessons of 19th century physics in the period of transition from Newtonian mechanics to field theory (section \ref{HertzS}.)  

\vspace{6pt}

I would like to thank Willem De Myunck, David Griffiths, Ehtibar Dzhafarov,  and~Marian Kupczynski for stimulating discussions and comments. This work was supported by the research project of the Faculty of Technology, Linnaeus University, 
``Modeling of complex hierarchic structures''.

\section*{Appendix 1: Incompatibility as Necessary Condition for Violation of Any Bell Type Inequality}\label{appa}

Consider a family of quantum observables $D_1,...,D_n$  represented by Hermitian operators 
$\hat D_1,..., \hat D_n.$ We restrict considerations to observables with discrete values;
thus, operators have the purely discrete spectra. Denote by $\hat E_i(x)$ the orthogonal projector 
corresponding to the eigenvalue $x$ of $\hat D_i.$    

Suppose that the observables are pairwise compatible, i.e.,~ 
$(D_i, D_j)$ can be measured jointly for any quantum state $\rho$ and  jpd is well defined
\begin{equation}
\label{JPD1}
p_{D_i D_j}(x,y; \rho)\equiv P(D_i=x, D_j=y; \rho).
\end{equation}

In QM, compatibility is mathematically represented via commutativity of operators, i.e.,~
$[\hat D_i, \hat D_j]=0,$ and, hence,  $[\hat E_i(x), \hat E_j(y)]=0.$ The quantum formalism gives the following formula for jpd
(von Neumann~\cite{VNM}):
\begin{equation}
\label{JPD2}
p_{D_i D_j}(x,y; \rho) =\rm{Tr} \; \rho E_i(x) E_j(y)=  \rm{Tr}\; \rho E_j(y) E_i(x) .
\end{equation}

Now, we point to the really surprising feature of quantum measurement theory. If~observables are pairwise 
compatible, i.e.,~each pair can be jointly measured with corresponding  
jpds $p_{ij}(x,y; \rho)$ given by Equation~(\ref{JPD2}), then they are also triple-wise compatible, quadruple-wise 
compatible and so on... Any family of observables, $D_{i_1},..., D_{i_m}$ 
can be jointly measured and the joint probability distribution is given by the formula:
\begin{equation}
\label{JPD3}
p_{D_{i_1} ... D_{i_m}}(x_1,...,x_m; \rho) =\rm{Tr} \; \rho E_{i_1}(x_1)... E_{i_m}(x_m).
\end{equation}

On the left-hand side of this formula, one can take any permutation of indexes.
This implies: 

\medskip

 ${\bf 2\implies m}:$ {\it pairwise compatibility $\implies$ multiple compatibility.} 

\medskip

This is really astonishing. It is surprising that its specialty (from the general viewpoint of measurement theory) was not discussed in foundational~literature.

We turn to Bell's inequalities. Now,
we are endowed with ${\bf 2\implies m}$ property of quantum~observables.  


Consider the most general Bell-type framework. There are $K$ groups of quantum observables:
$$
D^{k}=(D_1^{k},..., D_{N_k}^{k}),  k=1,..., K.
$$

Mathematically, they are represented by Hermitian operators:
$$
\hat D^{k}=(\hat  D_1^{k},...,\hat  D_{N_k}^{k}).
$$

Suppose that, for~different $k,$  observables  are compatible, i.e.,~in the mathematical framework:
$$
[\hat D_i^{n}, \hat  D_j^{m}]=0, n \not=m.
$$ 

Thus, jpds $p_{i_1...i_K}\equiv p_{D_{i_1}^{1}... D_{i_K}^{K}}$ are well defined and, hence,
covariations as well
$$
\langle D_{i_1}^{1}\cdots D_{i_K}^{K} \rangle = \rm{Tr} \; \rho \hat D_{i_1}^{1} \cdots \hat D_{i_K}^{K}=
\sum x_1 \cdots x_K \; p_{i_1...i_K}(x_1,...,x_K).
$$

Consider some Bell-type inequality (e.g., the~CHSH  inequality or the Mermin inequality),
\begin{equation}
\label{QTE}
\sum_{i_1...i_K} \;    t_{i_1...i_K} \;\langle D_{i_1}^{1}... D_{i_K}^{K} \rangle + \mbox{correlations of lower orders}  \leq c,
\end{equation}
where $t_{i_1...i_K}$ are some real constants. This inequality may be violated only 
if at least one pair  of observables, say $(D_{i}^{n}, D_{j}^{n}),$  is incompatible,
i.e., in~the mathematical formalism
\begin{equation}
\label{QTE1}
[\hat D_{i}^{n}, \hat D_{j}^{n}] \not=0.
\end{equation}

Otherwise, jpd exists and the inequality in Equation~(\ref{QTE}) cannot be~violated. 

\medskip

{\bf Theorem 2}
{\it Incompatibility is a necessary condition for violation 
of any Bell-type inequality.}


In the standard nonlocality discussions, it is assumed that there are $K$ systems $S^{k}, 
k=1,2,...,K,$ and observables $D^{k}$ are local observables for $S^{k}$. 
Endowed with this  scheme, we analyze the possibility to violate the Bell-type inequality in
Equation~(\ref{QTE}). The~necessary condition is that  Equation~(\ref{QTE1}) holds true.  
This condition is~local. 

\section{Appendix 2:``Entanglement'' in the Absence of the Tensor Product Structure}\label{appb}

In Section~\ref{EV}, we consider   compound systems and discuss  the well known fact that  eigenvectors of 
operator $\hat {\cal B}^2$ giving the $\max$-value of its quadratic form can be selected as separable 
(non-entangled) states; they need not be eigenvectors of the Bell operator; its eigenvectors are linear 
combinations of the aforementioned separable~states.

We want to show that the tensor product structure of states and operators is not crucial for the above 
consideration. 

Consider any Hermitian operator $\hat C$ and its square $\hat C^2.$ Let $u$ be an eigenvector
of the latter, i.e.,~$\hat C^2 u = \lambda u, \lambda >0,$ and let $u$ is not 
an eigenvector of the former. 
Set $v= \hat C u/\sqrt{\lambda},$ i.e.,
\begin{equation}
\label{Q2} 
\hat C  u= \sqrt{\lambda} v, \; 
\hat C v=  \sqrt{\lambda} u.
\end{equation}

Set
\begin{equation}
\label{Q21} 
\psi_{\pm}= u \pm v.
\end{equation} 

Then, $\hat C  \psi_{\pm}= \pm \sqrt{\lambda} \psi_{\pm}.$
Thus, $\psi_{\pm}$ are eigenvectors of $\hat C.$

If the quadratic form of $\hat C^2$ approaches its $\max$-value on eigenstate $u/\Vert u\Vert,$ then 
the quadratic form of $\hat C$ approaches its $\max$-value on eigenstate $\phi_{\pm}= \psi_{\pm}/ \Vert \psi_{\pm}\Vert .$

States $\psi_{\pm}$ can be considered as {\it generalization of entangled states}, which is to say entangled 
with respect to operator $\hat C.$

This consideration can be coupled to  measurement theory. Consider some quantum observable $D$ represented by Hermitian operator $\hat D.$ (For simplicity, suppose that $\hat D\geq 0.$) Suppose that this observable 
is simple theoretically, by~complex experimentally. (Spectrum and eigenvectors of operator $\hat D$ can be easily found, but~measurement of observable  $D$ is really challenging.) Consider also the observable $C$ represented by Hermitian operator $\hat C\equiv \sqrt{\hat D}.$ 
Suppose that this observable is complex theoretically, but~rather simple experimentally.
(The structure of  spectrum and eigenvectors of operator $\hat C$  is complicated,   but~ 
measurement design for $C$  is straightforward.)

We are interested in the following problem: {\it  Find experimentally the upper bound for  the average $\langle D \rangle_\psi$ of observable $D$ with respect to all possible states.} We stress that we are interested 
in the experimental verification of a theoretical prediction of~QM. 

We can easily find state $u$ corresponding to $\max$-eigenvalue $\lambda$ of operator $\hat D.$ Then, 
one of the $\max$-states of operator $\hat C$ can be found with the aid of ``$C$-entanglement'':
\begin{equation}
\label{H}
\phi_{+}= (u + \lambda^{-1/2} \hat C u)/ \Vert u + \lambda^{-1/2} Cu\Vert.
\end{equation}

Finally, we prepare an ensemble of systems in quantum state $\phi_+$ 
and perform $C$-measurement  for these~systems. 

In the Bell-type scenario (for observables respecting the tensor product structure), $\hat C=\hat {\cal B}$ is the Bell operator,  $\hat D=\hat {\cal B}^2.$
In fact, the~degree of incompatibility is encoded in the observable corresponding to  
operator $\hat D.$ However, its straightforward measurement would involve measurement of observables
corresponding to commutators. The~latter is challenging. At~the same time, eigenstates of $D$ have the 
simple tensor product structure (separable states). They can easily be found. Then,   eigenstates of the Bell 
operator can be generated as superpositions of Equation~(\ref{H}).

\section*{Appendix 3: Prequantum Classical Statistical Field Theory}

 The basic variables of PCSFT~\cite{PQ} are classical random fields defined on physical space.
A random field can be considered as a function of two variables $\phi= \phi(x; \omega)$: $x$ is the spatial variable \mbox{(with three} real coordinates); $\omega$ is a random parameter.
We remark that random fields can be considered as random vectors valued in the complex Hilbert space 
$H=L_2 ({\bf R}^3)$ of square integrable complex valued~functions. 

The key point of this theory is that covariance operator $B$ of random field $\phi$ is identified (up to normalization by trace) with the density operator of QM:
\begin{equation}
\label{LIK} 
B \to \rho=B/\rm{Tr} B.
\end{equation}

The covariance operator is an element of the descriptive theory (PCSFT) and the density operator is the element of the observational theory (QM). (For a complex valued random field, its covariance operator $B$ is a Hermitian positive operator with the finite trace. Thus, $B$ has all features of a density operator, besides~normalization $\rm{Tr} \rho=1.$)

We remark that here the trace of field's covariance operators equals 
to average of field's energy:
\begin{equation}
\label{LIK1} 
\rm{Tr} B= E \Vert \phi(\omega)\Vert^2,
\end{equation}
where $E$ is mathematical expectation and 
$$
\Vert \phi(\omega)\Vert^2 = \int_{{\bf R}^3} \vert \phi(x; \omega)\vert^2\; d x$$
is square of the $L_2$-norm of the field (for the concrete value of the random parameter $\omega).$   
Thus, normalization (determining ``descriptive $\to$ observational'' correspondence)  is with respect to field's~energy. 

Physical variables of PCSFT are quadratic forms of fields. Each quadratic form on $H=L_2$ is determined by a Hermitian operator, $\hat A: H\to H.$ Hence,  PCSFT variables have the form, 
$$f_A(\phi)(\omega) = \langle \phi(\omega)\vert A\vert \phi(\omega)\rangle,$$ 
where $\phi(\omega)\equiv \phi(x;\omega) \in L_2$ for each $\omega.$    
Quadratic forms are elements of the descriptive theory (PCSFT) and Hermitian operators are elements of the observational~theory. 

Averages calculated in PCSFT coincide with averages calculated in QM. However, the~range of values of a
quadratic form does not coincide with the range of values of the corresponding quantum observable, Hermitian
operator (cf. with descriptive theories of the Bell type).


\begin{thebibliography}{999}
\bibitem{EPR} Einstein, A.; Podolsky, B.; Rosen, N.   Can quantum-mechanical description of physical reality be considered complete? \emph{Phys. Rev.} {\bf 1935}, {\it 47}, 777--780.
\bibitem{BR0} Bohr, N.    \emph{The Philosophical Writings of Niels Bohr}; 
Ox Bow Press: Woodbridge, UK, 1987.
\bibitem{PL1} Plotnitsky, A.  \emph{Epistemology and Probability: Bohr, Heisenberg, Schr\"odinger and the Nature of   Quantum-Theoretical Thinking}; Springer: Berlin, Germany; New York, NY, USA, 2009.
\bibitem{PL2} Plotnitsky, A.  {\it Niels Bohr and Complementarity: An Introduction}; Springer: Berlin, Germany; New York, NY, USA, 2012.
\bibitem{BR} Bohr, N.  Can quantum-mechanical description of physical  reality be considered complete? \emph{Phys. Rev.} {\bf 1935},
 {\it 48},  696--702.
\bibitem{Bell0} Bell, J.S. On the Einstein Podolsky Rosen paradox. {\it Physics} {\bf 1964}, {\it 1}, 195--200.
\bibitem{Bell1} Bell, J.S. \textit{Speakable and Unspeakable in Quantum Mechanics}, 2nd ed.;  Cambridge University Press: Cambridge, UK, 2004.  
\bibitem{Bell2} Bell,  J.S.  On the problem of hidden variables in quantum theory. 
{\it Rev. Mod. Phys.} {\bf 1966}, {\it 38}, 450.

\bibitem{CHSH} Clauser, J.F.;  Horne, M.A.; Shimony, A.; Holt, R.A. Proposed experiment to test local hidden-variable theories. 
{\it Phys. Rev. Lett.} {\bf 1969}, {\it  23}, 880.


\bibitem{Accardi} Accardi, L. The probabilistic roots of the quantum mechanicalparadoxes. In {\it The Wave–Particle Dualism. \mbox{A Tribute} to Louis de Broglie on his 90th Birthday};  Diner S., Fargue D., Lochak G., Selleri F., Eds.;  D. Reidel Publication Company: Dordrecht, The Netherlands, 1984;
pp. 47–55.
\bibitem{KHBa} Khrennikov, A. {\it Interpretations of Probability}; VSP Int. Sc. Publishers: Utrecht/Tokyo, Japan, 1999; 2nd ed. (completed), De Gruyter: Berlin, Germany, 2009.

\bibitem{V2}  Khrennikov, A.  V\"axj\"o interpretation-2003: Realism of contexts. 
In  \emph{Quantum Theory: Reconsideration of Foundations}; Khrennikov A., Ed.; V\"axj\"o Univ. Press: V\"axj\"o, 2004, pp. 323--338.

\bibitem{[27ab]} Khrennikov, A. The principle of supplementarity: A contextual probabilistic viewpoint to complementarity, the interference of probabilities, and the incompatibility of variables in quantum mechanics.  \emph{Found. Phys.} {\bf 2005}, {\it  35}, 1655--1693.

\bibitem{KQL2}  Khrennikov, A.  Schr\"odinger dynamics as the Hilbert space projection of a realistic contextual probabilistic dynamics.   \emph{Europhys. Lett.} {\bf 2005}, {\it 69}, 678--684.

\bibitem{KC5} Khrennikov, A.  Bell-Boole inequality: Nonlocality or probabilistic incompatibility of random variables?
 \emph{Entropy} {\bf 2008}, {\it 10}, 19--32.

\bibitem{KHBb}  Khrennikov, A. {\it Contextual Approach to Quantum Formalism}; Springer: Berlin, Germany; New York, NY, \mbox{USA, 2009.}
\bibitem{KHB2} Khrennikov, A. Bohr against Bell: Complementarity versus nonlocality.  {\it Open Phys.} {\bf 2017},  {\it 15}, 734--738. 
\bibitem{KHB3} Khrennikov, A.;  Alodjants, A. Classical (local and contextual) probability model for Bohm–Bell type experiments: No-Signaling as independence of random variables. {\it Entropy} {\bf 2019}, {\it 21}, 157--177.
\bibitem{Muynck} De Muynck, W. {\it Foundations of Quantum Mechanics, an Empiricist Approach}; Springer: Dordrecht, \mbox{The Netherlands}, 2006.
\bibitem{Accardi1} Accardi, L. Some loopholes to save quantum nonlocality. In {\it Foundations of Probability and Physics-3}; AIP:
Melville, NY, USA, 2005; pp. 1–20, doi:10.1063/1.1874552.
\bibitem{Theo} Nieuwenhuizen, T.M.  Is the contextuality loophole fatal for the derivation of Bell inequalities?
{\it Found. Phys.} {\bf 2011}, {\it 41},  580--591.
\bibitem{DZ} Dzhafarov, E.N.; Kujala, J.V. Context-content systems of random variables: The contextuality-by default
theory. {\it J. Math. Psych.} {\bf 2016}, {\it 74}, 11--33.
\bibitem{BC1} Kupczynski, M. Can Einstein with Bohr debate on quantum mechanics be closed? \textit{Phil. Trans. Royal Soc. A}  
\textbf{2017},  \textit{375,}  2016039.
\bibitem{BC2} Kupczynski, M. Closing the door on quantum nonlocality. {\it Entropy} {\bf  2018}, {\it 20}, 877.
\bibitem{Griffiths} Griffiths, R.B.  Quantum nonlocality: Myth and reality. \emph{arXiv} \textbf{2019}, 
arXiv:1901.07050.
\bibitem{Boughn1} Boughn, S. Making sense of Bell's theorem and quantum nonlocality. {\it Found. Phys.} {\bf 2017}, {\it 47}, 640--657.
\bibitem{Boughn2} Boughn, S. There is no action at a distance in quantum mechanics, spooky or otherwise. \emph{arXiv} {\bf 2018},
arXiv:1806.07925.

\bibitem{Landau} Landau, L.J.  Experimental tests of general quantum theories.  \emph{Lett.  Math. Phys.} \textbf{1987}, \emph{14}, 33--40.

\bibitem{Landau1} Landau, L.J.  On the violation of Bell's inequality in quantum theory. {\it Phys. Lett. A}  {\bf 1987},  {\it 120}, 54--56.

\bibitem{Schilpp} Schilpp,  P.A. (Ed.) {Remarks to the Essays Appearing in this Collective Volume}. In {\it Albert Einstein: Philosopher-Scientist};  Tudor: New York, NY, USA,  1949; pp. 663--688. 
\bibitem{PL3} Plotnitsky, A.  {\it The Principles of Quantum Theory}; Springer: Berlin, Germany; New York, NY, USA, 2016. 


\bibitem{Braunstein} Braunstein,  S.L.; Mann,  A.;  Revzen, M. 
Maximal violation of Bell inequalities for mixed states. {\it \mbox{Phys. Rev. Lett.}} {\bf 1992}, {\it 68}, 3259--3261.
\bibitem{Hardy} Hardy, L. $N$-measurement Bell inequalities, N-atom entangled states, and the nonlocality of one photon. {\it Phys. Lett. A}  {\bf 1991}, {\it 160},  1. 
\bibitem{Cereceda}  Cereceda, J.L.   Maximally entangled states and the Bell inequality. {\it Phys. Lett. A}  {\bf 1996}, {\it 212},  123--129. 
\bibitem{Mermin} Mermin, N.D. Extreme quantum entanglement in a superposition of macroscopically distinct states.  {\it \mbox{Phys. Rev. Lett.}} {\bf 1990}, {\it 65}, 1838--1840.   

\bibitem{Wolf}  Wolf, M.M.;  Perez-Garcia, D.; Fernandez,  C.    
Measurements incompatible in quantum theory cannot be measured jointly in any other local theory.
{\it Phys. Rev. Lett.} {\bf 2009}, {\it 103}, 230402.



\bibitem{BS} Bogoliubov,  N.N.; Shirkov,  N.N.   {\it Introduction to Theory of Quantized Fields}; 
 Interscience Publishers: Moscow, Russia, 1959.
\bibitem{Haag} Haag, R.  {\it Local Quantum Physics. Fields, Particles, Algebras}; Springer: Berlin/Heidelberg, Germany, 1996.
\bibitem{NQFT} Greiner, W.;   Reinhardt, J.  {\it Field Quantization}, Part II; 
Springer: Berlin/Heidelberg, Germany, 1996. 
\bibitem{Shimony} Shimony, A. Hidden-variables models of quantum mechanics (Noncontextual and contextual). \mbox{In {\it Compendium
of Quantum Physics};} Springer: Berlin/Heidelberg, Germany, 2009; pp. 287--291.
\bibitem{Shimony1} Shimony, A. Experimental test of local hidden variable theories. In {\it Foundations of Quantum Mechanics};
Academic: New York, NY, USA, 1971.
\bibitem{Jaeger} Jaeger, G. {\it Quantum Information: An Overview}; Springer: Berlin, Germany; New York, NY, USA, 2007. 
\bibitem{Jaeger1} Jaeger, G. {\it Quantum Objects: Non-Local Correlation, Causality and Objective Indefiniteness in the Quantum World};
Springer: Berlin, Germany; New York, NY, USA, 2013.

\bibitem{Hertz1} Hertz, H.  {\it The Principles of Mechanics: Presented in a New Form}; Macmillan: London, UK, 1899.
\bibitem{Boltzmann} Boltzmann, L.  Uber die Frage nach der objektiven Existenz der 
Vorgnge in der unbelebten Natur. In {\it  Populre 
Schriften}; Barth, J.A., Ed.;  Vieweg+Teubner Verlag: Leipzig, Germany, 1905. 
\bibitem{Boltzmann1} Boltzmann, L.  On the development of the methods of 
theoretical physics in recent times.  
In {\it Theoretical Physics and Philosophical Problems}; McGuinness, B., Ed.; Springer: Dordrecht, \mbox{The Netherlands}, 1974; Volume~5, Vienna Circle 
Collection.
\bibitem{DA} D’Agostino, S.  Continuity and completeness in physical 
theory: Schr\``odinger’s return to the wave interpretation of quantum 
mechanics in the 1950's. In  
{\it E. Schr\``odinger: Philosophy and the Birth of Quantum Mechanics}; Bitbol, M., Darrigol, O., Eds.; 
 Editions Frontieres: Gif-sur-Yvette, France, 1992; pp. 339--360.
\bibitem{Hertz} Khrennikov, A. Hertz's viewpoint on quantum theory. {\it Act. Nerv. Super.}  \textbf{2019}, \emph{61}, 24--30,
doi:10.1007/s41470-019-00052-1.
\bibitem{PQ} Khrennikov, A.  \emph{Beyond Quantum}; Pan Stanford Publication: Singapore, 2014.
\bibitem{Aspect} Aspect, A.;  Dalibard, J.;  Roger, G.  Experimental test of Bell’s Inequalities using time-varying analyzers. {\it \mbox{Phys. Rev. Lett.}} {\bf 1982},  {\it 49}, 1804--1807.
\bibitem{Hensen} Hensen, B.; Bernien, H.; Dreau, A.E.; Reiserer, A.; Kalb, N.; Blok, M.S.; Ruitenberg, J.; Vermeulen, R.F.; Schouten, R.N.; Abellan, C.;~et~al. Experimental loophole-free violation of a Belli nequality using entangled 
electron spins separated by 1.3 km. {\it  Nature} {\bf 2015}, {\it 526}, 682.
\bibitem{Giustina}    Giustina, M.; Versteegh, M.A.; Wengerowsky, S.; H.; steiner, J.; Hochrainer, A.; Phelan, K.; Steinlechner, F.; Kofler, J.; Larsson, J.Å; Abellan, C.;~et~al. A significant-loophole-free test of Bell’s theo-rem with entangled photons.
{\it Phys. Rev. Lett.} {\bf 2015},  {\it 115}, 250401.
\bibitem{Shalm} Shalm, L.K.; Meyer-Scott, E.; Christensen, B.G.; Bierhorst, P.; Wayne, M.A.; Stevens, M.J.; Gerrits, T.; \mbox{Glancy, S.;} Hamel, D.R.; Allman, M.S.;~et~al. A strong loophole-free test of local realism. {\it Phys. Rev. Lett.} {\bf 2015}, {\it 115}, 250402.  
\bibitem{Aspect1} Aspect, A. Closing the door on Einstein and Bohr’s quantum debate. {\it Physics} {\bf 2015},  {\bf 8}, 123. 
\bibitem{Wiseman}  Wiseman, H. Quantum physics: Death by experiment for localrealism. \emph{Nature} \textbf{2015}, \emph{526}, 649--650.

\bibitem{ABELL} Khrennikov, A. After Bell. {\it Fortschritte der Physik (Progress in Physics)}.
Topical issue – International Conference Frontiers of Quantum and Mesoscopic Thermodynamics Prague, 
Czech Republic 27 July - 1 August 2015. {\bf 2017},  {\it 65},  N 6-8, 1600014.

\bibitem{Weihs} Weihs, G.; Jennewein, T.; Simon, C.; Weinfurther, H.; Zeilinger, A. Violation of Bell's Inequality under Strict Einstein Locality Conditions. \emph{Phys. Rev. Lett.} \textbf{1998}, \emph{81},
5039--5043.
\bibitem{VNM} Von Neumann, J. {\it Mathematical Foundations of Quantum Mechanics}; Princeton University Press: Princeton, NJ,
USA, 1955.

\end{thebibliography}
 \end{document}